\documentclass{elsarticle}
\usepackage{palatino}
\usepackage[USenglish]{babel}
\usepackage{times}
\usepackage{amssymb,amsmath}
\usepackage{graphics}
\usepackage{verbatim}
\usepackage[caption=false]{subfig}
\usepackage{url}
\usepackage{listings}
\usepackage[pdftex,colorlinks=true,linkcolor=blue,citecolor=blue,urlcolor=blue]{hyperref}
\usepackage{natbib}

\usepackage[usenames,dvipsnames]{color}
\usepackage[normalem]{ulem}

\begin{document}
\title{ESPResSo++ 2.0: Advanced methods for multiscale molecular simulation}
\author[mpip] {Horacio V. Guzman}
\author[mpip]{Nikita Tretyakov}
\author[mpip]{Hideki Kobayashi}
\author[mpip]{Aoife C. Fogarty}
\author[mpip]{Karsten Kreis}
\author[ul]{Jakub Krajniak}
\author[lanl]{Christoph Junghans}
\author[mpip]{Kurt Kremer}
\author[mpip]{Torsten Stuehn\corref{cor1}}
\ead{stuehn@mpip-mainz.mpg.de}

\address[mpip]{Max Planck Institute for Polymer Research, Ackermannweg 10, 55128 Mainz, Germany}
\address[ul]{KU Leuven Department of Computer Science, Celestijnenlaan 200A, 3001 Leuven, Belgium}
\address[lanl]{Computer, Computational, and
Statistical Sciences Division, Los Alamos National Laboratory, Los Alamos, NM 87545, USA}
\cortext[cor1]{Corresponding author}

\date{\today}

\begin{abstract}
Molecular simulation is a scientific tool used in many fields including material science and biology. This requires constant development and enhancement of algorithms within molecular simulation software packages. Here, we present computational tools for multiscale modeling developed and implemented within the ESPResSo++ package. These include the latest applications of the adaptive resolution scheme, the hydrodynamic interactions through a lattice Boltzmann solvent coupled to particle-based molecular dynamics, the implementation of the hierarchical strategy for equilibrating long-chained polymer melts and a heterogeneous spatial domain decomposition.

The software design of ESPResSo++ has kept its highly modular C++ kernel with a Python user interface. Moreover, it has been enhanced by automatic scripts that parse configurations from other established packages, providing scientists with the ability to rapidly set up their simulations.
\end{abstract}
\maketitle

\section{Introduction}
\label{intro}

 Molecular simulation methods~\cite{KK-FMP2002,attig04,voth08,holm0509,peter_faraday_2010,Halverson2013,OlveraSM2k15,RapaportCPC1991,ShawPCHPCNSA2009} have facilitated the study, exploration and co-design\cite{GuzmanBJON2017} of diverse materials. The functional and dynamic properties of biological and non-biological materials at diverse length and time scales can be simulated with sequential coarse-graining methods~\cite{KK-FMP2002,votca13} or with concurrent coarse-graining and atomistic methods~\cite{Donadio_PRL_2013-hadres_molliq,BD_US_AL_2007}. Such multiscale methods often involve coarse-graining the atomistic degrees of freedom into effective degrees of freedom representing a collection of atoms, entire monomers or even molecules~\cite{KK-FMP2002}. An important benefit of multiscale methods is to achieve computational speed-up, which comes from both the coarse-graining method itself and also from optimized algorithms~\cite{GuzmanPRE2017}.

Within the past decades numerous researchers have contributed to simulations packages like GROMACS~\cite{gromacs4}, LAMMPS~\cite{lammps95}, NAMD~\cite{namd2005}, ESPResSo~\cite{limbach06a}, ESPResSo++~\cite{Halverson2013}, among many others. These packages have been devoted to the development of Molecular Dynamics (MD) atomistic and coarse-graining simulations, giving rise to highly parallelizable and flexible codes. The latter, flexibility, is the main design goal of ESPResSo++. Thanks to the flexibility of the ESPResSo++ package, users can easily extend simulation methods in order to meet theoretically and experimentally driven goals, such as in the case of multiscale simulations~\cite{OlveraSM2k15}. In addition, ESPResSo++ combines flexibility and extensibility with the computational requirements of high-performance computing platforms via an MPI-based parallelization. One proof of ESPResSo++'s flexibility for multiscale simulations is the implementation and extension of the Adaptive Resolution Scheme (AdResS) to its Hamiltonian-based version (H-AdResS) and the addition of features such as the flexible spatial atomistic resolution regions approach\cite{Kreis2016c}.
ESPResSo++ can be easily used as a molecular dynamic engine and combined with other algorithms, for example, to study complex chemical reactions at the coarse-grained scale~\cite{DeBuyl2015,Krajniak2018}, or to do reverse mapping from a coarse-grained to an atomistic scale using an adaptive resolution approach~\cite{Krajniak2016, Krajniak2017}.

In this article we have selected two multiscale simulation methods that have been implemented in ESPResSo++, namely, concurrent multiple resolution simulations using AdResS~\cite{Praprotnik2008review,PraprotnikJDSKK2008,poblete_coupling_2010, praprotnik_statistical_2011} and the lattice Boltzmann technique which can be coupled to particle-based simulations\cite{PA_BD_1999}. 

The Adaptive Resolution Scheme has been used for simulations using diverse techniques, ranging from concurrent simulations of classical atomistic and coarse-grained models~\cite{praprotnik_adaptive_2005,praprotnik_adaptive_2007-1,praprotnik_corrigendum:_2009,Donadio_PRL_2013-hadres_molliq,Kremer_JCP_2015-adresprot,Kreis2015,Kreis2016b,Praprotnik_JCTC_2015-dna}, to interfacing classical atomistic with the path-integral formulation of quantum    models~\cite{Kreis2016,Kreis2017}, as well as interfacing particle-based simulations with continuum mechanics~\cite{delgado08a,delgado-buscalioni_coupling_2009}. Systems that have been simulated with the Adaptive Resolution Schemes implementation in the ESPResSo++ package include homogeneous fluids~\cite{praprotnik_adaptive_2005,praprotnik_adaptive_2007-1,praprotnik_corrigendum:_2009}, biomolecules in solution\cite{Kreis2014,Kremer_JCP_2015-adresprot,Kreis2015} and DNA molecules in salt solution~\cite{Praprotnik_JCTC_2015-dna}. In the present ESPResSo++ release, we come closer to the requirements of adaptive resolution schemes in terms of scalability. Here we also included the Heterogeneous Spatial Domain Decomposition Algorithm (HeSpaDDA), namely a density-aware spatial domain decomposition with moving domain boundaries, for AdResS and H-AdResS simulations, applicable to heterogeneous systems like nucleation, evaporation, and crystal growth.

The second simulation technique introduced in the present release of ESPResSo++ is the Lattice Boltzmann (LB) method, which accounts for hydrodynamic interactions in fluids and can be applied to many problems, ranging from studies of turbulence on the macroscale to soft matter investigations on the microscale. The latter includes hybrid simulations of particle-based systems, \textit{e.g.} colloids or polymers, in a solvent. The virtue of the LB method with respect to the explicit solvent treatment is its methodological locality and, as a consequence, computational efficiency. The LB-module of ESPResSo++ can be used: (i) as a stand-alone method for studies of turbulence and liquids driven by body-forces or (ii) in combination with molecular dynamics providing correct hydrodynamics (in contrast to, \textit{e.g.}, Langevin thermostat).

On top of the highlighted methods for multiscale molecular simulations available in ESPResSo++, this new release also introduces the implementation details of the hierarchical strategy for the equilibration of dense polymer melts. The hierarchical equilibration strategy comprises a recursive coarse-graining algorithm with its corresponding sequential back-mapping~\cite{zhang2014equilibration}.  

The contents of this publication are focused on the introduction of the new or updated methods and algorithms within the second release of ESPResSo++. The adaptive simulation schemes are described in Sec.~\ref{adress} while the Lattice Boltzmann method is presented in Sec.~\ref{latB}. The hierarchical strategy for the equilibration of polymer melts is described in Sec.~\ref{hierarchical}. In Sec.~\ref{mrDD} we present the deployment of the HeSpaDDA algorithm. Sec.~\ref{sec:devWF} provides information on how to contribute to the development of ESPResSo++. Finally, sec.~\ref{sec:intO} reports on the integration of ESPResSo++ with other useful packages.

Regarding the development of ESPResSo++, we want to highlight its user-friendly environment due to the Python interface used for the simulation scripts, and thus higher degrees of freedom to interact with other scientific software parts of the Python community, e.g. \texttt{NumPy}~\cite{numpy}, \texttt{SciPy}~\cite{scipy}, \texttt{scikit-learn}~\cite{PedregosaJMLR2011}, \texttt{Pandas}~\cite{pandas} and \texttt{PyEMMA}~\cite{PyEMMA}. Those wishing to get started with the package should visit our webpage~\cite{eppWEB} or directly go to our GitHub repository~\cite{eppGH}. Directions for downloading and building ESPResSo++ are given in both references. Finally for more details of the methods, algorithms or general code of ESPResSo++ please make use of our documentation~\cite{eppdoc} and previous publication~\cite{Halverson2013}.

\section{Adaptive resolution simulations}
\label{adress}
\subsection{Introduction}
\begin{figure}[!t]
\centering
\includegraphics[clip,width=0.65\columnwidth,keepaspectratio]{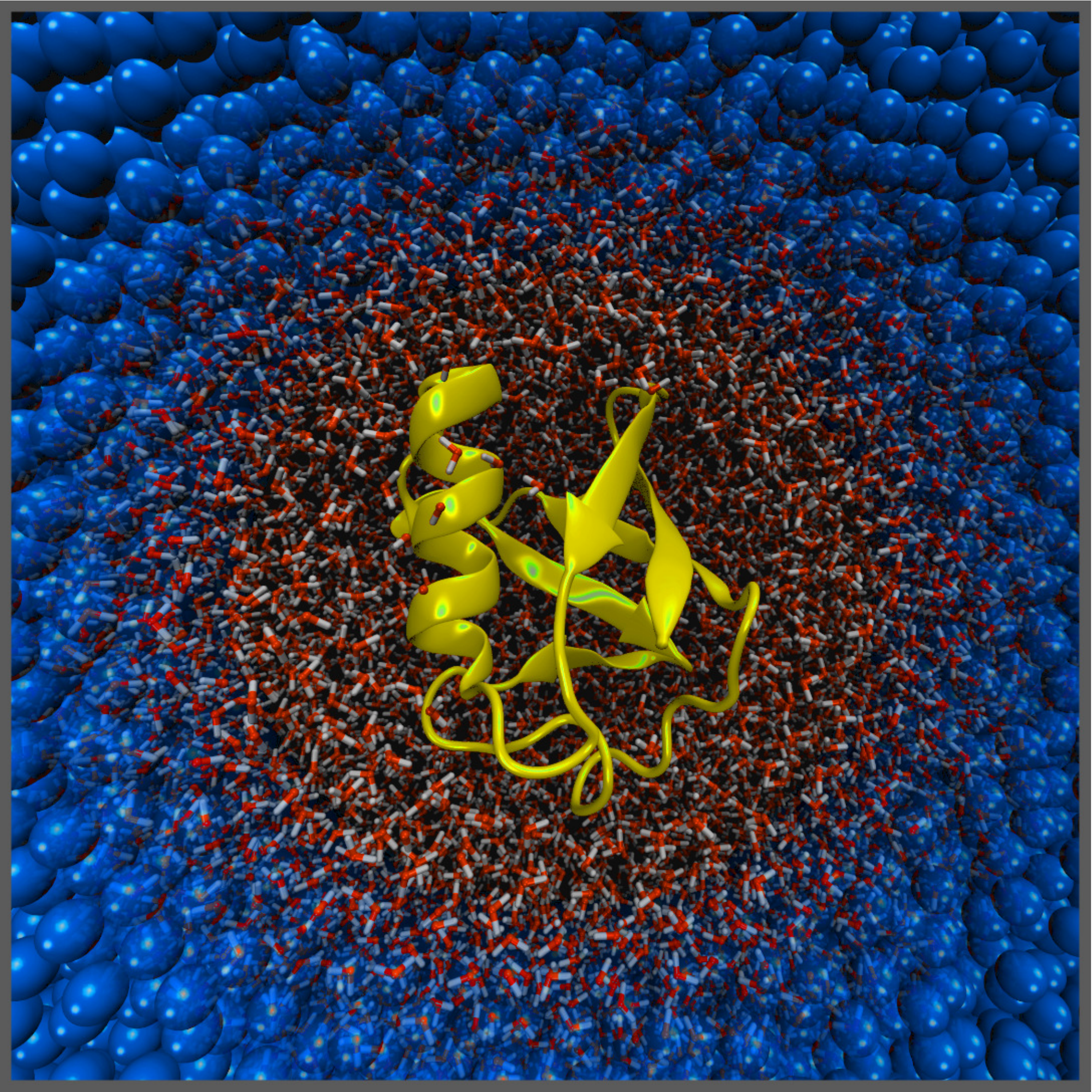}
\caption{AdResS simulation of an atomistic protein and its atomistic hydration shell, coupled to a coarse-grained particle reservoir via a transition region.\cite{Kremer_JCP_2015-adresprot}}
\label{fig: adress app}
\end{figure}

Heterogeneous systems containing a wide range of length- and timescales can be challenging to model using molecular simulation. This is because high-resolution, chemically detailed models are needed to describe certain processes or regions of interest; however, such models are also computationally expensive, and using them to model the entire system can be prohibitive. One approach to tackle this problem is the use of multi-resolution simulation techniques, in which more expensive, typically atomistic, and cheaper, typically coarse-grained models are used within the same simulation box, allowing one to reach longer overall length- and timescales \cite{Praprotnik2008review,HMM,Roehm2015,vanGunsteren_AngewChimie_2013-multisc_bio_review,DelleSite2017}. In such techniques, a region in space is defined in which molecules are modeled using atomistic detail, while coarse-grained models are used elsewhere (see examples in Figure~\ref{fig: adress app}). The Adaptive Resolution Simulation (AdResS) methodology deals with the coupling between atomistic (AT) and coarse-grained (CG) models \cite{Kremer_JCP_2005-onthefly,Praprotnik2008review,Donadio_PRL_2013-hadres_molliq,Kreis2014,Potestio2014a,Kreis2016b,DelleSite2017}. In this methodology, solvent particles can freely diffuse between AT and CG regions, smoothly changing their resolution as they cross a hybrid or transition region. The AdResS approach can be useful for modeling a wide variety of different systems, such as simple solutes in dilute solution and complex biomolecular systems \cite{Lambeth2010,Fritsch2012toluene,Mukherji2012,Mukherji2013,Wang2013gcadress,Praprotnik_JChemPhys_2014-adres_bio,Praprotnik_JCTC_2015-dna,Kremer_JCP_2015-adresprot,Kreis2015,Kreis2016c,Peters2016,Sablic2016,Netz2016,Fogarty2016}.

ESPResSo++ provides full support of the AdResS methodology, and the ESPResSo++ AdResS implementation has been used to simulate systems ranging from homogeneous fluids to biomolecules in solution \cite{Kreis2014,Kremer_JCP_2015-adresprot,Kreis2015,Kreis2016b,Kreis2016c,Fogarty2016,Kreis2017,Fogarty2017}. In the AdResS approach, the coupling of AT and CG models can take place via an interpolation on the level of either forces or energies. Force-interpolation AdResS was included in release 1.0 of ESPResSo++. Energy-interpolation (known as Hamiltonian- or H-AdResS), as well as the latest features are presented in this article. 

\subsection{Force interpolation}
In AdResS, a typically small part of the system, the AT region, is described on the AT level and coupled via a hybrid (HY) transition region to the CG region, where a coarser, computationally more efficient model is used. The interpolation is achieved via a resolution function $\lambda({\mathbf{R_\alpha}})$, a smooth function of the center of mass position $\mathbf{R_\alpha}$ of molecule $\alpha$. For each molecule, its instantaneous resolution value $\lambda_\alpha = \lambda({\mathbf{R_\alpha}})$ is calculated based on the distance of the molecule from the center of the AT region. It is 1 if the molecule resides within the AT region and smoothly changes via the HY region to 0 in the CG region (see, for example, Ref. \cite{Kreis2016b}). 

In the force interpolation scheme, the original AdResS technique \cite{Kremer_JCP_2005-onthefly,Praprotnik2008review}, two different non-bonded force fields are coupled as 
\begin{equation}\label{eq:adress}
\mathbf{F}_{\alpha|\beta} = \lambda(\mathbf{R_\alpha})\lambda(\mathbf{R_\beta})\mathbf{F}_{\alpha|\beta}^{\text{AT}} + \left(1-\lambda(\mathbf{R_\alpha})\lambda(\mathbf{R_\beta})\right)\mathbf{F}_{\alpha|\beta}^{\text{CG}},
\end{equation}
where $\mathbf{F}_{\alpha|\beta}$ is the total force between the molecules $\alpha$ and $\beta$ and $\mathbf{F}_{\alpha|\beta}^{\text{AT}}$ defines the atomistic force-field, which is decomposed into atomistic forces between the individual atoms of the molecules $\alpha$ and $\beta$. Finally, $\mathbf{F}_{\alpha|\beta}^{\text{CG}}$ is the CG force between the molecules, typically evaluated between their centers of mass.
Note that in addition to the non-bonded interactions usually also intramolecular bond and angle potentials are present. As these are computationally significantly easier to evaluate, they are typically not subject to any interpolation and therefore not further discussed here.

\subsection{Potential energy interpolation}
\label{sec:hadres}
Alternatively, the atomistic and the CG models can also be interpolated on the level of potential energies. In the Hamiltonian adaptive resolution simulation approach (H-AdResS) \cite{Donadio_PRL_2013-hadres_molliq,Potestio2013b,Espanol2015a}, the Hamiltonian of the overall system is defined as  
\begin{eqnarray}\label{Hmix}
H & = &\sum_\alpha \sum_{i\in\alpha} \frac{\mathbf{p}_{\alpha i}^2}{2m_{\alpha i}}+\sum_{\alpha} \left\{\lambda({\mathbf{R_\alpha}}) V^{\text{AT}}_\alpha + \left(1 - \lambda({\mathbf{R_\alpha}})\right) V^{\text{CG}}_\alpha \right\},
\end{eqnarray}
where $m_{\alpha i}$ and $\mathbf{p}_{\alpha i}$ are, respectively, the mass and the momentum of atom $i$ of molecule $\alpha$. The single-molecule potentials $V^{\text{AT}}_\alpha$ and $V^{\text{CG}}_\alpha$ are the sums of all non-bonded intermolecular interaction potentials corresponding to the AT and the CG model acting on molecule $\alpha$. We again omitted additional intramolecular interactions. To ensure a smooth transition of the molecules between the low- and high-resolution regions, one typically chooses a non-linear interpolation function $\lambda$, such as a squared cosine \cite{Kreis2016c,Kreis2016b}. Furthermore, the CG potential is often parametrized to have an excluded volume similar to the AT potential. This prevents overlapping particles and diverging forces when molecules travel between the regions. Alternatively, a force-capping mechanism can be applied if necessary \cite{Kreis2015}. The intramolecular atomistic degrees of freedom, which are subject to bonded and angular interactions, are either frozen when a molecule resides in the CG region or they are simply integrated everywhere in the full system including the CG region, since they are typically computationally very efficient to evaluate.

The crucial difference between the force-based and the potential energy-based adaptive resolution scheme is an additional force term, dubbed drift force, arising in the forces corresponding to the Hamiltonian in Eq. \ref{Hmix} (for details, see \cite{Donadio_PRL_2013-hadres_molliq}). Both schemes have characteristic advantages and disadvantages and which approach is better suited for implementing the adaptive coupling depends on the application. On the one hand, the force interpolation technique exactly preserves Newton's third law, but it does not allow a Hamiltonian formulation, does not conserve the energy \cite{DelleSite2007}, and it therefore requires thermostatting for stable simulations \cite{Praprotnik2006a,DelleSite2007,praprotnik_adaptive_2007,praprotnik_fractional_2007,Poblete2010,Kremer_Ensing_comment}. The method can be used for the calculation of expectations in the canonical ensemble, which relies on thermostating anyway, and even for the estimation of dynamical quantities in the AT region when only the CG region is thermostated to dissipate the excess energy \cite{Kremer_JCP_2015-adresprot,Kreis2016c}. 

H-AdResS, on the other hand, also allows microcanonical simulations and other approaches that require the existence of a well-defined Hamiltonian. However, it violates Newton's third law in the hybrid region and it features an additional undesired force that must be explicitly taken care of. H-AdResS can be applied when the exact dynamics, i.e. the exact preservation of Newton's third law, is only relevant in the AT region. Importantly, the additional force term that stems from the application of the position-derivative on the resolution function $\lambda({\mathbf{R_\alpha}})$ in Eq. \ref{Hmix} and that reflects the Helmholtz free energy difference between the AT and the CG models is only a minor hindrance. Its effect, pushing particles from one region to the other, can be efficiently canceled out on average using so-called free energy corrections (also see Sec. \ref{sec: fec tf}). The H-AdResS methodology is particularly advantageous when one is interested in energy-conserving calculations or other simulations that rely on a Hamiltonian formulation, such as path integral-based techniques (see Sec. \ref{PI-AdResS}) \cite{Donadio_PRL_2013-hadres_molliq,Kreis2016,Kreis2017}.

Both adaptive resolution schemes have been successfully applied for simulations of various different systems, including complex liquids, solutes in dilute solutions, large biomolecules, polymers and quantum systems \cite{Lambeth2010,Fritsch2012toluene,Potestio2012parahydrogen,Mukherji2012,Mukherji2013,Wang2013gcadress,Praprotnik_JChemPhys_2014-adres_bio,poma,Poma2011,Praprotnik_JCTC_2015-dna,Kremer_JCP_2015-adresprot,Kreis2015,Kreis2016c,Peters2016,Sablic2016,Netz2016,Fogarty2016,Kreis2016,Kreis2017}. The increase in computational efficiency compared to fully high-resolution simulations varies and depends strongly on the size and properties of the system, the employed AT and CG models, as well as the parallelization and domain-decomposition scheme (also see Sec. \ref{mrDD}). In previous applications, the intermolecular force calculations, the part of the simulation that is actually modified in AdResS, were speeded up by factor of up to $\approx 9$, while overall reported simulation speed-ups are of the order $\approx$ 2-3 \cite{Kreis2015,Kremer_JCP_2015-adresprot,Kreis2016,Kreis2016c,Kreis2017}.

\subsection{Free energy corrections and the thermodynamic force}
\label{sec: fec tf}
Typical CG models have significantly different pressures compared to the AT reference systems~\cite{Stillinger2002,Louis2002,Wang2009,DAdamo2013}. In adaptive resolution simulations, this leads to a pressure gradient between the AT and the CG subsystems, which, in addition to the drift force in H-AdResS, pushes particles across the HY transition region.

Therefore, a correction field must be applied in the HY region to enforce a flat density profile along the direction of resolution change. This compensation force counteracts the pressure gradient and cancels the drift force in H-AdResS. An appropriate correction can be derived, for example, via Kirkwood thermodynamic integration \cite{Donadio_PRL_2013-hadres_molliq,Kirkwood1935}. This is known as free energy correction and in particular useful in H-AdResS. An alternative approach, frequently used in the force interpolation method, is to construct a correction force directly from the distorted density profile obtained without any correction and then refine it in an iterative fashion. This approach is known as the thermodynamic force \cite{ciccotti_adress}.

ESPResSo++ allows the straightforward inclusion of such correction forces and also includes routines that can be used to calculate density profiles, pressures and energies, required for deriving these corrections.

\subsection{Self-adjusting adaptive resolution simulations}
Many complex systems, such as proteins, membranes and interfaces, do not feature regular spherical or planar geometries. Furthermore, they undergo large-scale conformational changes during simulation. Therefore, recently a scheme was derived that, within the framework of forced-based AdResS, allows AT regions of any arbitrary shape. Additionally, the AT region can change its geometry during the simulation to follow, for example, a folding peptide \cite{Kreis2016c}. This is established by associating several spherical AT regions with many atoms of a macromolecule, such that their overlap defines an envelope around the extended object. When it deforms, this shell adapts accordingly. 

This scheme is available in ESPResSo++ within the force interpolation approach and in combination with the thermodynamic force.

\subsection{Multiple time stepping in adaptive resolution simulations}
Since CG potentials are typically significantly softer than AT force fields, the corresponding equations of motions can be solved using a larger time step. This suggests the use of multiple time stepping (MTS) techniques in adaptive resolution simulations, in which both AT and CG potentials are present simultaneously. A RESPA-based MTS approach \cite{Tuckerman1992,TuckermanBook} is now available in ESPResSo++, which enables different time steps for updates of the CG and the AT forces.

\subsection{Thermodynamic integration}
As explained above in Sec.~\ref{sec:hadres}, no global Hamiltonian can be defined in the force-interpolation version of AdResS. Nevertheless, the potential-energy-based Thermodynamic Integration (TI) approach to free energy calculations can be combined with force-interpolation AdResS, as recently shown using simulations of amino acid solvation in ESPResSo++.\cite{Fogarty2017} This is because AdResS allows the sampling of local configurations which are equivalent to those of fully atomistic simulations.

The TI implementation in ESPResSo++ can also be used to perform standard fully atomistic free energy calculations, such as to calculate solvation free energies or ligand binding affinities, among other examples.

\subsection{Path integral-based adaptive resolution simulations}\label{PI-AdResS}
The path integral (PI) formalism can be used in molecular simulations to account for the quantum mechanical delocalization of light nuclei~\cite{Feynman1965,TuckermanBook}. It is frequently used, for example, when modeling hydrogen-rich chemical and biological systems, such as proteins or DNA~\cite{Perez2010,Li2010,Nagata2012,Pamuk2012,Wang2014}. In the PI methodology, quantum particles are mapped onto classical ring polymers, which represent delocalized wave functions. This renders the PI approach computationally highly expensive (for a detailed introduction see, for example, \cite{TuckermanBook}). 

In practice, the quantum mechanical description is often only necessary in a small subregion of the overall simulation. Recently, a PI-based adaptive resolution scheme was developed that allows to include the PI description only locally and to use efficient classical Newtonian mechanics in the rest of the system~\cite{Kreis2016,Kreis2017}. In this approach, the ring polymers are forced to collapse to classical, point-like particles in the classical region via a mechanism similar to the potential energy interpolation outlined above. In the PI formalism the strength of the spring constants of the ring polymers representing the quantum particles depends on the particles' masses. A light particle has weaker springs between the beads of the ring polymers, resulting in more extended ring polymers, which correspond to more delocalized particles. Similarly, heavy particles have very stiff springs leading to more collapsed ring polymers corresponding to more classical behavior. In the PI-AdResS scheme, one works in normal mode coordinates and interpolates between light and much heavier masses for the degrees of freedom that correspond to the ring polymers' internal modes and ring vibrations, using a resolution function just like in the other adaptive resolution schemes described previously. In that way, the ring polymers behave quantum-mechanically in the high-resolution region and effectively classically in the low-resolution region. Importantly, when the ring polymers are collapsed to point-like particles in the classical, low-resolution subregion, the potential energy calculation becomes significantly more efficient, since one does not need to consider the different contributions from different ring polymer beads anymore.

The method is based on an overall Hamiltonian description and it is consistent with a bottom-up PI quantization procedure. It allows for the calculation of both quantum statistical as well as approximate quantum dynamical quantities in the quantum subregion using ring polymer or centroid molecular dynamics. The methodology is implemented in the ESPResSo++ package and it also makes use of multiple time stepping. It has been used for the calculation of the structural and dynamical properties of quantum-mechanically modeled water as well as liquid hydrogen. For a more technical discussion of the scheme see Refs. \cite{Kreis2016,Kreis2017}.

\section{Lattice Boltzmann}
\label{latB}
    
\subsubsection*{Introduction}
The Lattice Boltzmann (LB) method in ESPResSo++ was designed for efficient simulations of phase-separating semidiluted polymer solutions. These solutions are characterized by: (i) a low volume fraction $\phi < 5\%$ of polymeric material with respect to the system's volume $V$ and (ii) long polymer chains that start to overlap. These requirements are satisfied for spatially large systems with only a few very long chains. Since it is not computationally feasible to treat the solvent as explicit particles (their number would be much greater than several millions) we rely on the lattice-based LB methodology in the solvent treatment~\cite{RB_SS_1992,YQ_DD_1992,SS_2001,BD_AL_2009,CA_JC_2010}. The polymer chains are modeled by molecular dynamics (MD).

\begin{figure}
  \centering
  \subfloat[]{\label{fig:goodsolvent}
    \includegraphics[width=0.32\textwidth]{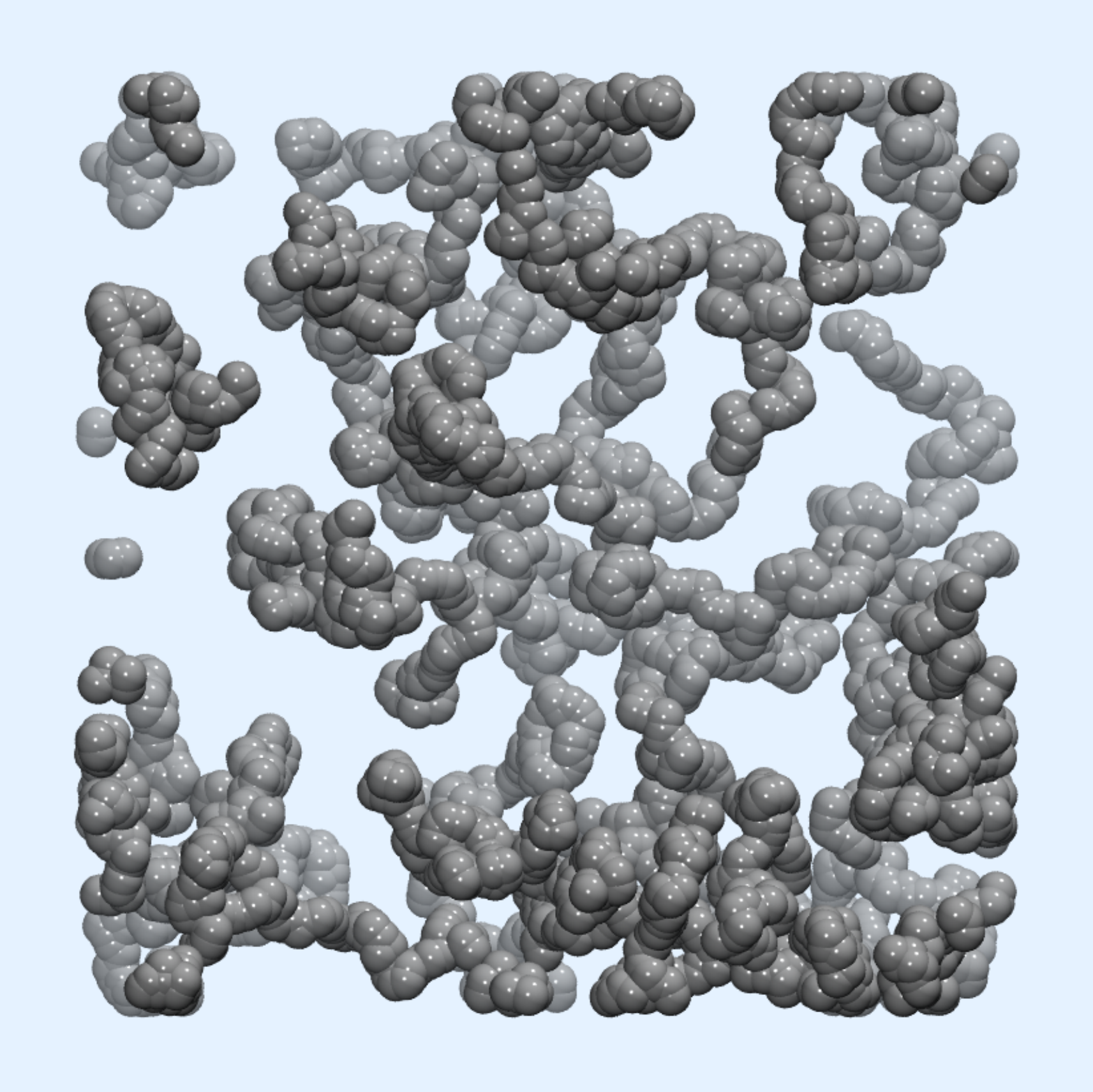}}\hspace*{0.2cm}
  \subfloat[]{\label{fig:poorsolvent}\includegraphics[width=0.32\textwidth]{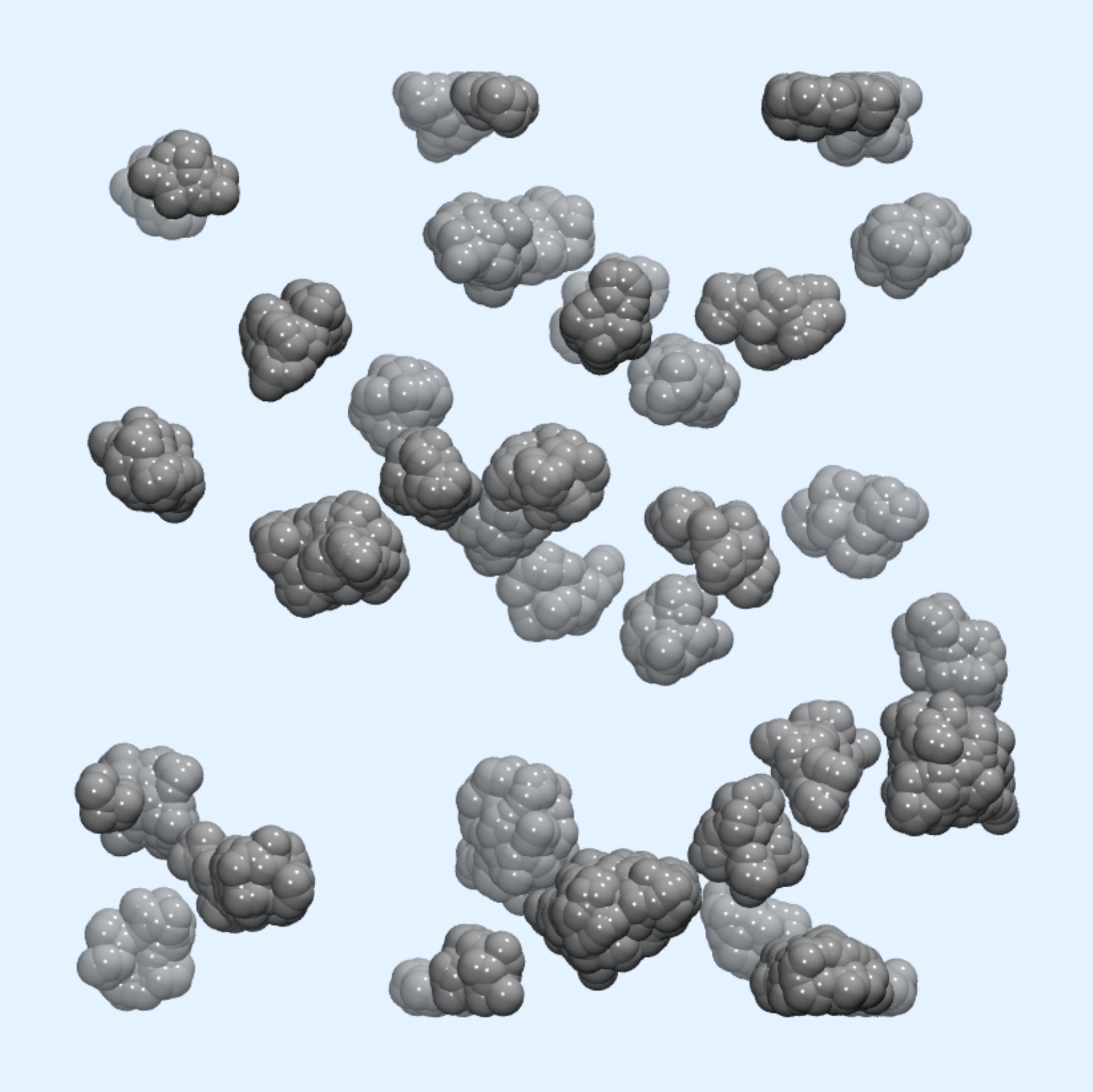}}\hspace*{0.2cm}
  \subfloat[]{\label{fig:coarsening}\includegraphics[width=0.32\textwidth]{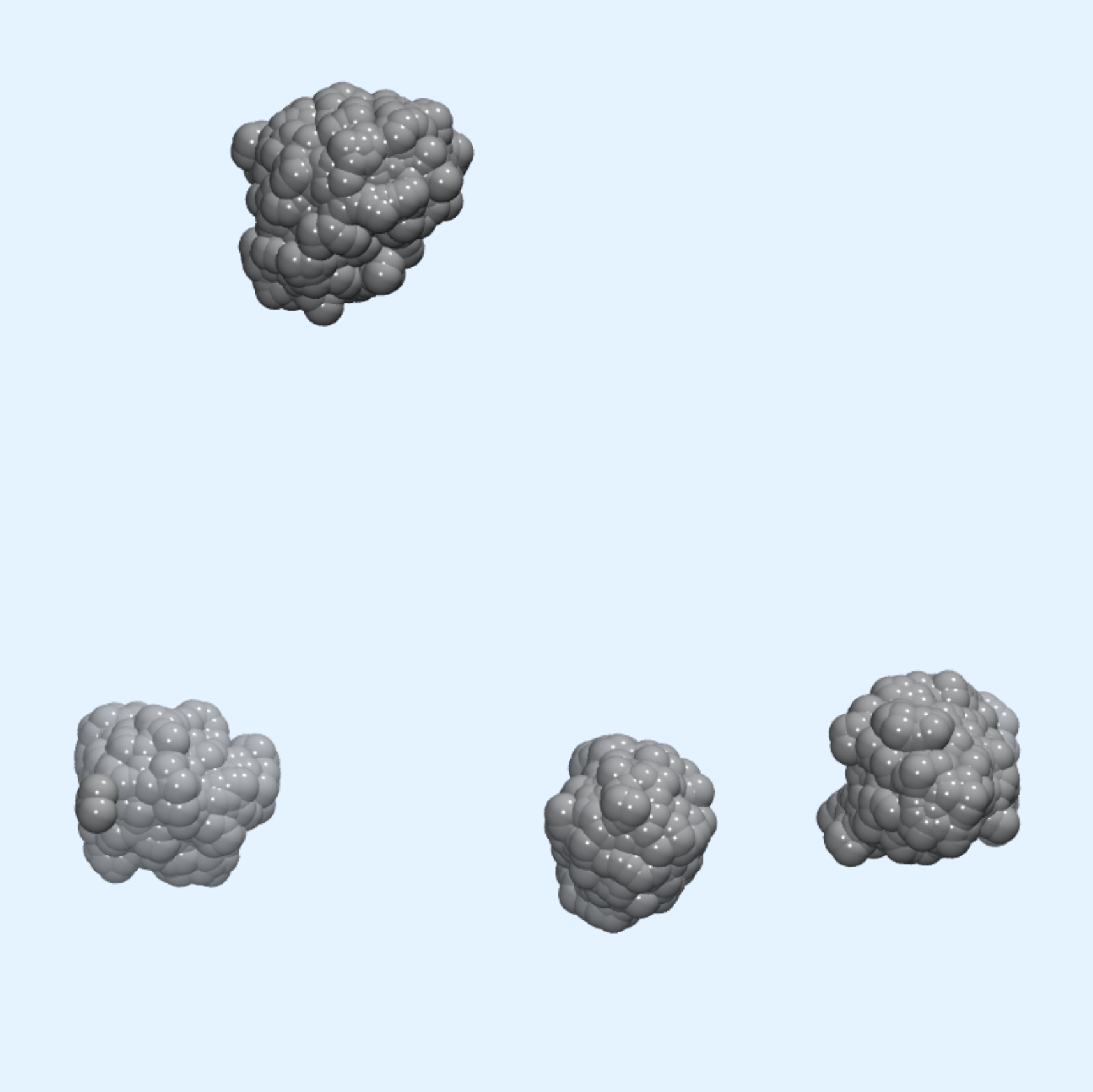}}
  \caption{(a) Good solvent, (b) Poor solvent and (c) Coarsening.}
  \label{fig:LB-MD-coupling}
\end{figure}

The hybrid LB/MD method is used to study the phase-separation of the polymer solution upon the change of the solvent quality. Under good solvent conditions the chains are extended coils, as the interactions between their monomers and the solvent are favorable. This situation is shown in Fig.~\ref{fig:goodsolvent}. A quench into poor solvent regime (Fig.~\ref{fig:poorsolvent}) initiates the collapse of the polymers and sets out a slow coarsening (Fig.~\ref{fig:coarsening}), i.e. agglomeration of individually collapsed chains into multichain polymeric droplets. The quenched system evolves on multiple time and length scales and demonstrates rich dynamical properties. 

\subsubsection*{Implementation details}
The LB technique can be viewed as a version of coarse-graining of the solvent fluid on a lattice. At every lattice site $\vec{r}$ and time $t$ the fluid is modeled by a set of single-particle distribution functions or populations $f_i(\vec{r},t)$. The sites are connected in one LB timestep, $\Delta t_{\textrm{LB}}$, by the finite set of velocities, $\vec{c_i}$. Mass and momentum density are given by $\rho = \sum_i f_i$ and $\vec{j_i} = \sum_i f_i \vec{c_i}$, respectively . In ESPResSo++ we employ a popular three-dimensional D3Q19 model with $19$ velocity vectors $\vec{c_i}$~\cite{YQ_DD_1992}.

The LB step is divided into collision and streaming parts. At first, the populations collide according to the kinetic rules given by a collision operator. As the operator we use the multiple-relaxation times scheme~\cite{DdH_IG_MK_2002} that allows a straightforward introduction of thermal fluctuations~\cite{BD_US_AL_2007} relevant to soft matter research. In the streaming phase the post-collisional populations are propagated to the neighboring sites according to velocity vectors $\vec{c_i}$ and the LB step is finished.

\begin{figure}
  \centering
  \subfloat[]{\label{fig:LB_MD_coupling}\includegraphics[width=0.53\textwidth]{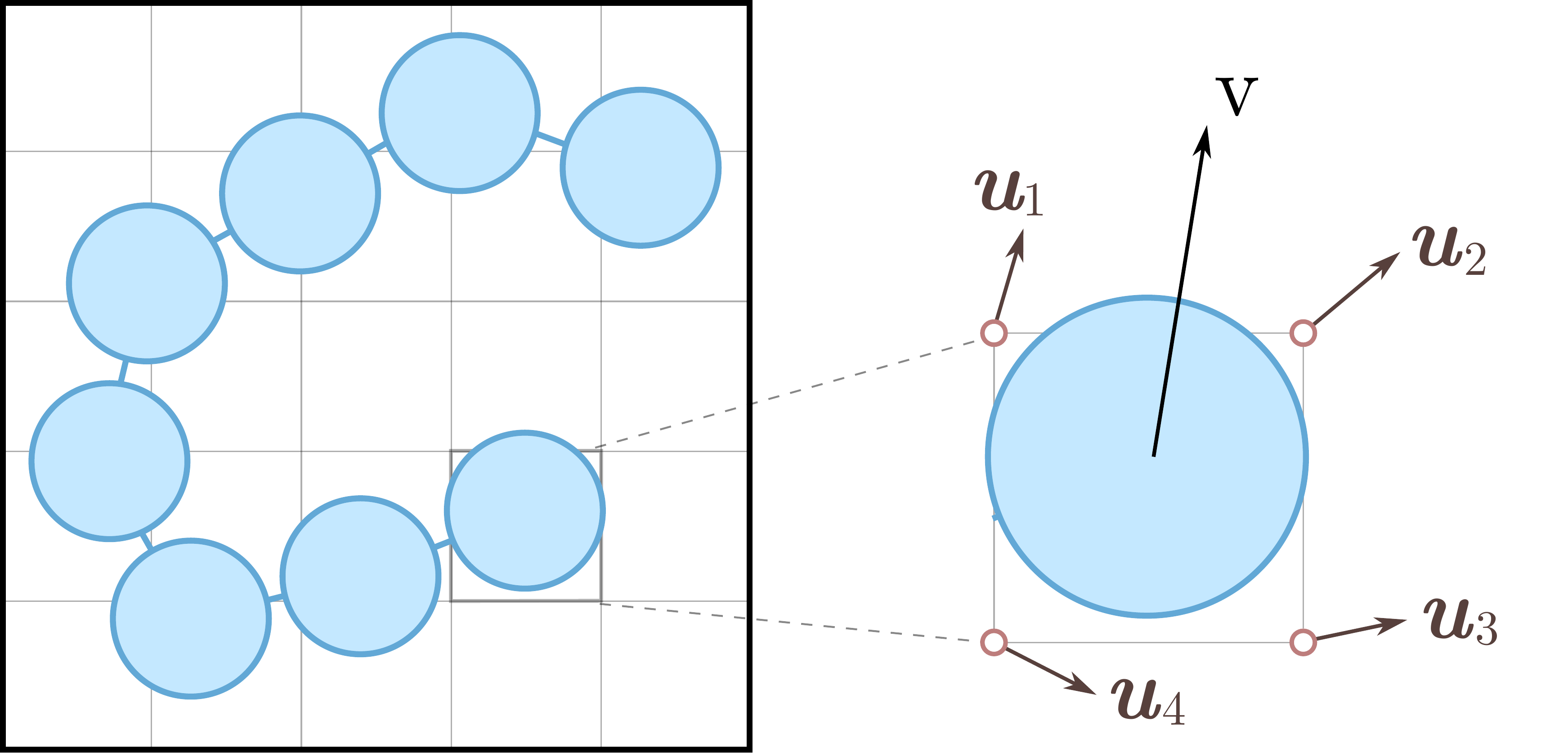}}\hspace*{0.3cm}
  \subfloat[]{\label{fig:LB_MD_timescale}
    \includegraphics[width=0.38\textwidth]{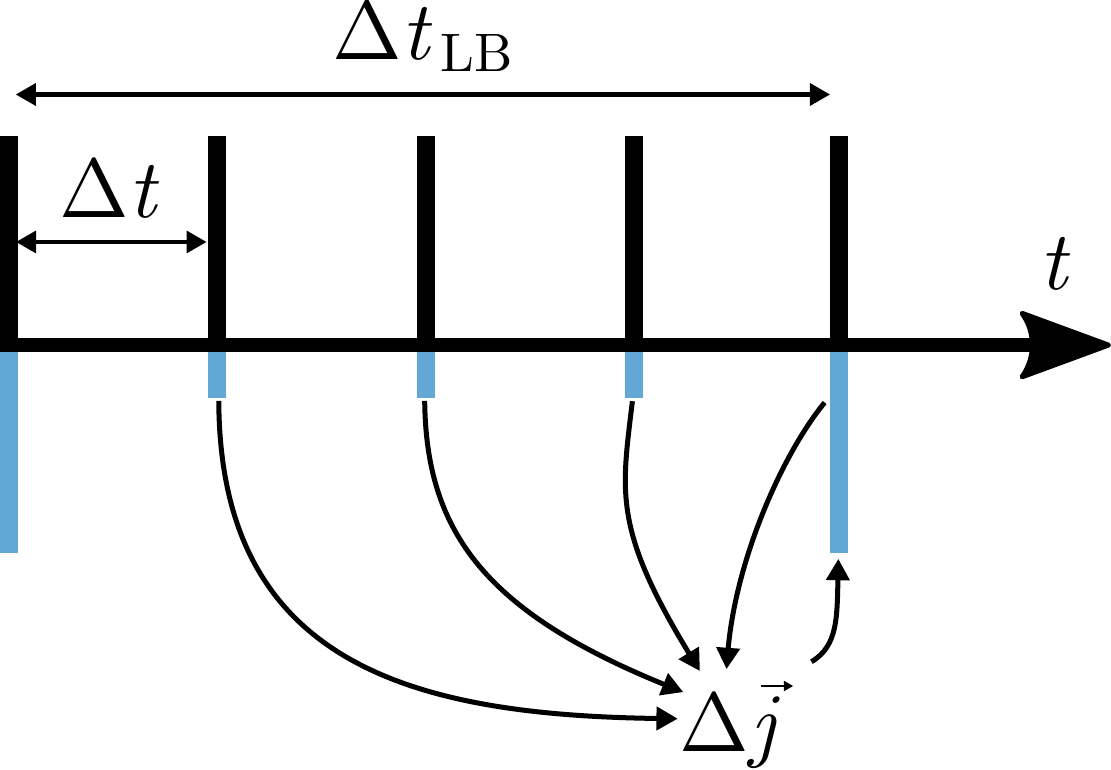}}
  \caption{(a) Schematic representation of the LB-to-MD coupling. The two-dimensional plot is used for simplification. The implemented method is three-dimensional. (b) Time scales separation between the LB and MD time scales.  }
\end{figure}

The coupling between the LB fluid and MD particles is done in a dissipative fashion~\cite{PA_BD_1999} as sketched in Fig.~\ref{fig:LB_MD_coupling}. The force $\vec{F}$ exerted by a solvent onto an MD particle located at position $\vec{R}$ and moving with velocity $\vec{\textnormal{v}}$, is given by
$\vec{F} = - \zeta [ \vec{\textnormal{v}} - \vec{u}(\vec{R})] + \vec{F}_\mathrm{rand}$,
where $\vec{F}_\mathrm{rand}$ is the random force due to thermal motion, and the first term is the viscous friction with an amplitude $\zeta$. This term accounts for the velocity of the MD particle $\vec{R}$ with respect to the velocity of the fluid at the position of the particle $\vec{u}(\vec{R})$. The latter is interpolated from the fluid velocities $\vec{u_i}$ at the neighboring lattice sites.

To conserve total momentum of the LB/MD system a counterforce $-\vec{F}$ should act from the MD particle onto the LB fluid. We recast this force in terms of the momentum change $-\vec{F} = \Delta \vec{j} / \Delta t$,
%
where $\Delta t$ is the MD timestep. The momentum change $\Delta \vec{j}$ of the LB fluid is distributed to the neighboring lattice sites. 
  
A time-costly LB step is done only after several MD steps~\cite{PA_BD_1999}, so we use $\Delta t_\textrm{LB} / \Delta t = 5$ or $10$. In this approach, the forces $\vec{F}$ onto MD particles are calculated in every MD step, while the concommitant momentum changes $\Delta \vec{j}$ at the LB sites are accumulated in memory. Firstly, they update the fluid velocities in every MD step: $\vec{u}_i \rightarrow \vec{u}_i + \Delta \vec{j}_i / \rho_i$, where $\rho_i$ and $\vec{j}_i$ are the mass and momentum density of the LB fluid at the site $i$. Secondly, the accumulated momentum changes are applied at the LB collision step via correction of the collision operator.  This algorithm is shown in Fig.\ref{fig:LB_MD_timescale}. For a detailed description of the method we address the reader to Ref.~\cite{TretyakovCPC2017}.

\subsubsection*{Efficiency}

The LB method of ESPResSo++ employs a regular lattice. Along with the extreme locality of the algorithm (only neighboring sites are connected) it profits from a straightforward but efficient parallelisation strategy realised by MPI (message-passing interface). The hybrid LB/MD approach conserves hydrodynamics and is more feasible for large simulations than explicit solvent treatment. Moreover, the timestep separation between MD and LB realised in ESPResSo++ facilitates further speed-up, as a time-intensive LB update is done only every several MD steps.

\section{Hierarchical equilibration strategy for polymer melts}
\label{hierarchical}
\subsection*{Introduction}

To study the properties of polymer melts by numerical simulations,
we have to prepare equilibrated configurations. However,
the relaxation time for polymer melts increases, according to reptation
theory, with the third power of the molecular weight \cite{de1971reptation,doi1978dynamics,doi1988theory}. In fact, equilibrated
configurations of high molecular weight polymer melts cannot be
obtained by brute-force calculation in a realistic time, i.e. the
CPU time for $1000$ polymers consisting of $2000$ monomers is roughly
estimated as about $4.0\times10^6$ hours on a single processor ($2.2$GHz) on the basis of reptation
theory \cite{de1971reptation,doi1978dynamics,doi1988theory} and actual measured CPU time per one particle per step.
Hence, an effective method for decreasing the
equilibration time is required. The hierarchical equilibration strategy pioneered in
Ref. \cite{zhang2014equilibration, zhang2015communication} is a particularly suitable way to do
this.

The hierarchical equilibration strategy consists of recursive coarse-graining and sequential back-mapping \cite{zhang2014equilibration}.
At first, a polymer chain, originally consisting
of $M$ monomers, is replaced by a coarse-grained (CG) chain consisting
of $M/N_{b}$ softblobs, mapping from each subchain with $N_{b}$ monomers, represented as the model developed by Vettorel \cite{vettorel2010fluctuating}.
In this model, the relaxation time doesn't increase in accordance with the
reptation theory but rather Rouse theory, since the CG chains can pass through each other. The degree of freedom of the system is $N_{b}$ times less than that of the microscopic model.
Hence, the relaxation time of CG chain configuration is drastically decreased.
After equilibrating a configuration at a very coarse resolution, each CG
polymer chain is replaced with a more fine-grained (FG) chain. In this
back-mapping procedure, a CG blob is divided into several FG blobs. The
center of mass (COM) of the FG blobs coincides with the position of the
CG blob's center and is kept fixed during the relaxation of the
local conformation of the FG monomers within the CG blob.
Consequently, a microscopic equilibrated configuration can be reproduced
by sequential back-mapping.

The required functions of this strategy have been implemented into ESPResSo++.  

\begin{figure}[!t]
\centering
\includegraphics[clip,width=0.9\columnwidth,keepaspectratio]{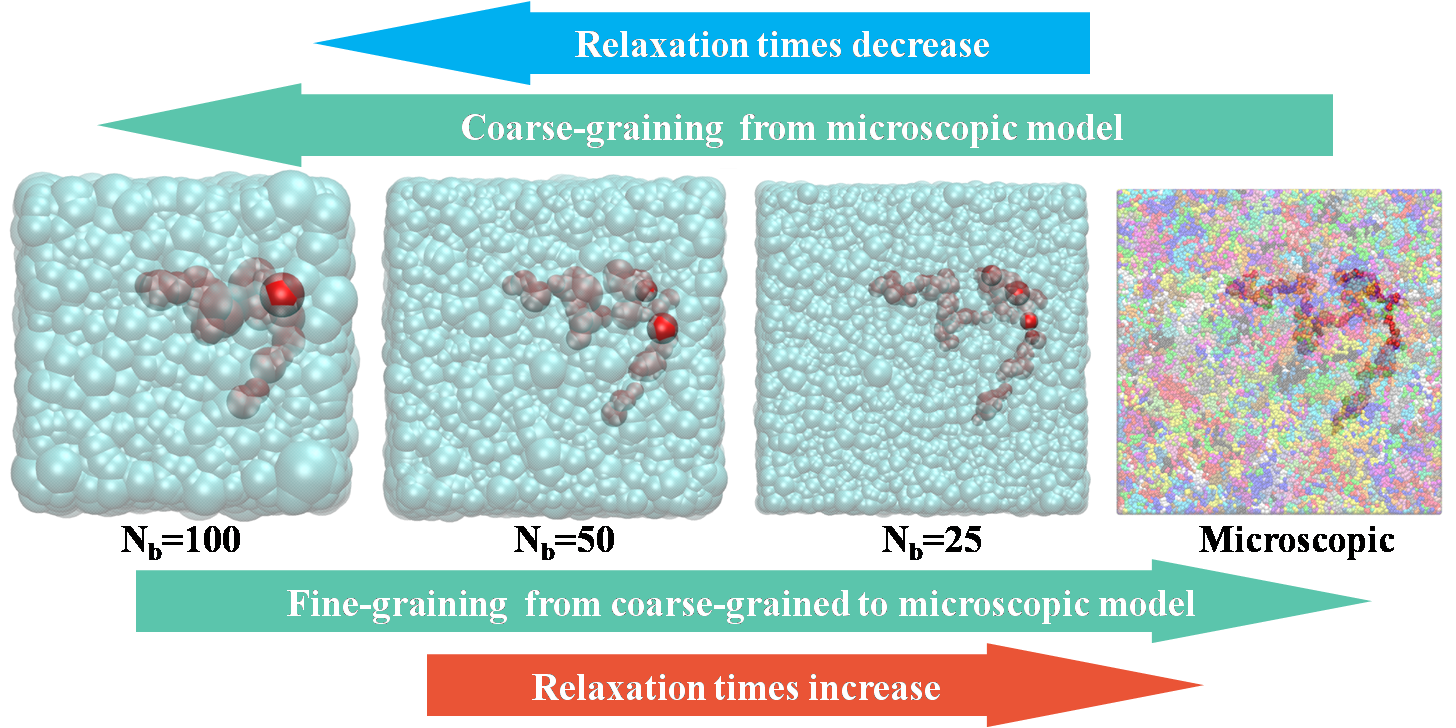}
\caption{The schematic representation of hierarchical equilibration
 strategy. To decrease relaxation time, the microscopic configurations are
 mapped to a coarse-grained soft blob model. After equilibrating
 configurations at a very coarse resolution, the microscopic resolution is reproduced
 by sequential back-mapping.}
\label{fig: hierarchical_irst}
\end{figure}

\begin{figure}[!t]
 \centering
 \includegraphics[clip,width=0.9\columnwidth,keepaspectratio]{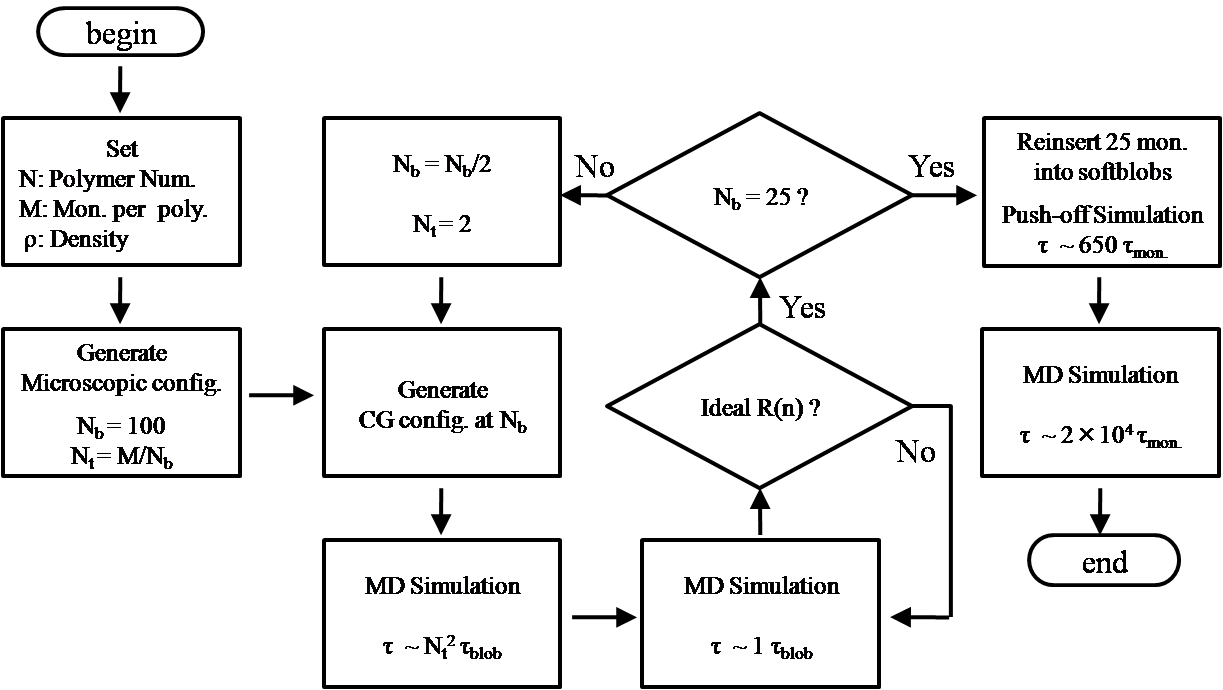}
 \caption{The flowchart of the hierarchical equilibration
 strategy}
 \label{fig: flowchart_hierarchical}
\end{figure}

\begin{figure}[tb]
\centering
 \subfloat[]{\includegraphics[clip,width=0.45\columnwidth,keepaspectratio]{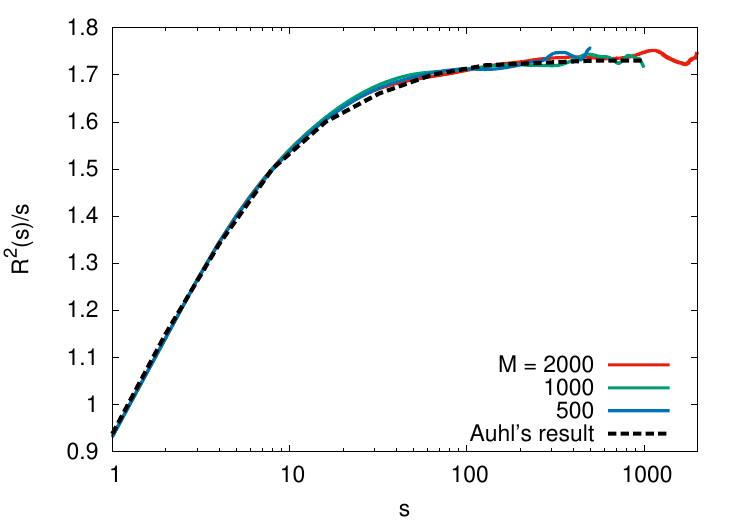}\label{fig: msid_n1_various}}
 \subfloat[]{\includegraphics[clip,width=0.45\columnwidth,keepaspectratio] {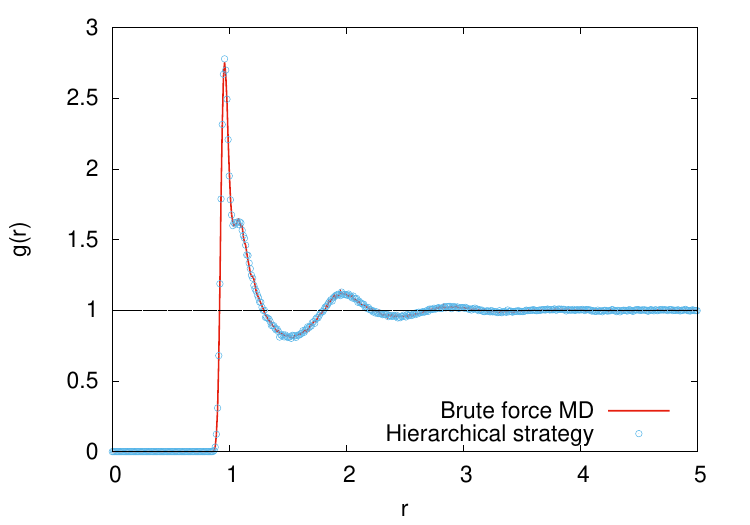}\label{fig: pcf_compare}}
 \caption{(a) The mean square internal distances with various polymerisation degree M. M = 500 (blue), 1000 (green) and 2000 (red).  The dashed line stands for Auhl's result \cite{auhl2003equilibration}. (b) Pair correlation function g(r) with the polymerisation degree M=500 taken from the hierarchical equilibration strategy (symbol) and the brute force molecular dynamics simulations.}
 \label{fig: demonstration_equilibrated_system}
\end{figure}

To demonstrate that the system is equilibrated, we evaluate the mean square internal distance (MSID), defined as $\langle R_{ij}/s_{ij} \rangle$ where $R_{ij}$ is the distance between $i$ and $j$-th monomer on a chain and $s_{ij}$ is the contour length between two monomers. The MSID is the slowest physical property to equilibrate. Hence, other properties already have their equilibrated value, when the MSID reaches the equilibrated form. Auhl's work \cite{auhl2003equilibration} already obtained the equilibrated MSID form. We use their result as a reference MSID. Fig.~\ref{fig: msid_n1_various} shows the MSID equilibrated by the hierarchical strategy for $M=500$, $1000$ and $2000$. All of them converge to the same line as Auhl's result \cite{auhl2003equilibration}. For confirmation, we also present the pair correlation function $g(r)$ for $M=500$ obtained from brute force calculation and our hierarchical strategy shown in Fig.~\ref{fig: pcf_compare}. They are in quite good agreement with each other. Thus, we conclude that our hierarchical strategy can fully equilibrate the polymer melts system.

 \subsubsection*{Efficiency}
 
 The efficiency of the hierarchical strategy for polymer melts can be
 estimated by a comparison with the brute-force calculation.
 The CPU time for brute-force calculations $\tau_{\rm{brute}}$ is
 described as
 $\tau_{\rm{brute}} \sim N \times M^3 \times \left(M/N_{e}
					    \right) \tau_{\rm{mon}}$,
 where $N_{e}$ stands for the number of monomers between entanglement. This value is obtained for the product of the number of monomers, $N \times M$, and the reptation time, $M^2 \times (M/N_{e}) \tau_{\rm{mon}}$, \cite{doi1978dynamics}.

 The computational time for the hierarchical strategy
 $\tau_{\rm{hier}}$ is defined as the summation of the computational time at
 various resolutions. Thus, we should estimate the relaxation time at various resolutions.
 The relaxation time $\tau_{100}$ for the softblob at $N_{b}=100$ can be estimated as
 $\tau_{100} \sim N \times \left(M/100\right)^3 \tau_{\rm{blob}}$.
 The relaxation time $\tau_{50}$ for the softblob at $N_{b}=50$ is roughly estimated as
 $\tau_{50} \sim N \times (M/50) \times 64 \tau_{\rm{blob}}$.
 In a similar way, the relaxation time $\tau_{25}$ for the softblob at
 $N_{b}=25$ can be estimated as 
 $\tau_{25} \sim N \times (M/25) \times 64 \tau_{\rm{blob}}$.
 Practically, the calculation time $\tau_{\rm{push}}$ for the push-off
 procedure and $\tau_{\rm{micro}}$ for the microscopic model have been
 defined as $650NM\tau_{\rm{mon}}$ and about $2NM\times10^{4}\tau_{mon}$
 respectively \cite{moreira2015direct}.
 At all MD simulations except the push-off procedure, the increment time
 $dt$ is defined as the unit time divided by $200$. Only at the push-off
 procedure, $dt$ is defined as the unit time divided by $10^4$. Thus,
 $\tau_{\rm{push}}$ should be multiplied by $50$ when we estimate the
 computational effort.
 Additionally, the cpu time for a step per particle per processor ($2.2$GHz) is about
 $4.0\times10^{-7}$ seconds for the microscopic model and about $4.0\times10^{-5}$ seconds
 for softblob models. Thus, $\tau_{\rm{blob}}$ should be multiplied by
 $100$ for estimating the efficiency. Hence, the computational time
 $\tau_{\rm{hier}}$ is estimated as
 $\tau_{\rm{hier}} \sim
   100 \times \tau_{100} +javascript:void(0);
   100 \times \tau_{50} +
   100 \times \tau_{25} +
    50 \times \tau_{\rm{push}} +
    \tau_{\rm{micro}}$.
 
 As a consequence, we can estimate the efficiency of the hierarchical
 strategy by the ratio of $\tau_{\rm{hier}}$ and $\tau_{\rm{brute}}$ represented as
 \begin{equation}\label{eq:efficiency_large_M}
   \frac{\tau_{\rm{hier}}}{\tau_{\rm{brute}}} \sim 10^{-4}\frac{N_{e}}{M}+4.2884 \times 10^{4}\frac{N_{e}}{M^{3}}.
 \end{equation}
 For example, after substituting $M=2000$ and $N_{e}=100$ to $\tau_{\rm{hier}}/\tau_{\rm{brute}}$, we obtain the concrete value of the ratio
  \begin{equation}\label{eq:efficiency}
   \frac{\tau_{\rm{hier}}}{\tau_{\rm{brute}}} \approx \frac{1}{1.9\times10^3}.
 \end{equation}
 Please note that the efficiency of the hierarchical strategy increases with increasing $M$.

\section{Heterogeneous Spatial Domain Decomposition Algorithm}
\label{mrDD}
\subsubsection*{Introduction}
\begin{figure}[thb!]
\centering
\includegraphics[clip,width=1.0\columnwidth,keepaspectratio]{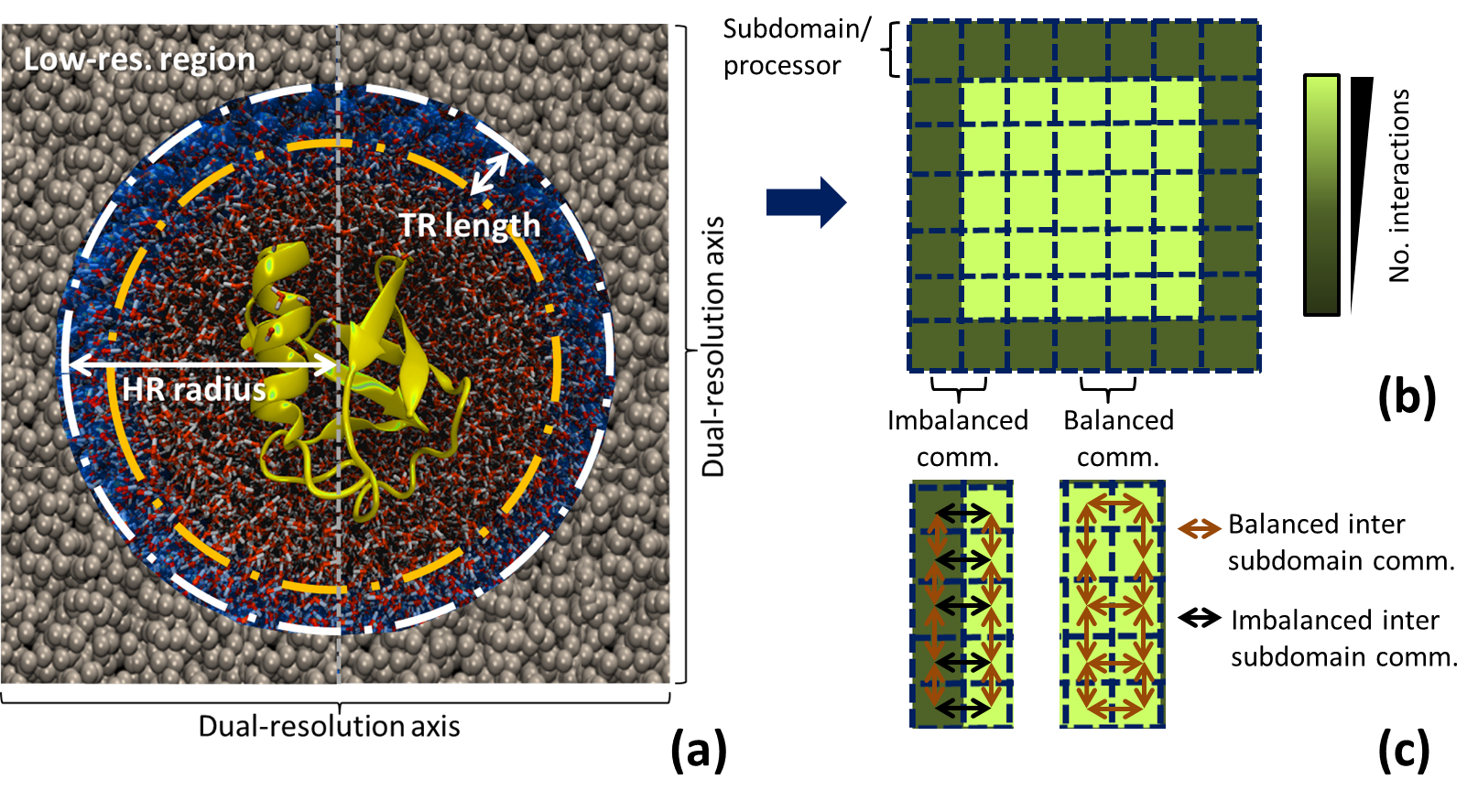}
\caption{AdResS simulation of an atomistic protein and its atomistic hydration shell, coupled to a coarse-grained particle reservoir via a transition region~\cite{Kremer_JCP_2015-adresprot}. (a) Illustrates all details of the multiscale system subregions. The low-resolution is in gray while the high-resolution is marked by its radius and a white circle. The transition or hybrid region is also marked between the white and orange circles. (b) Scheme of interactions load (computationally exhaustive) as for the subdomains homogeneous distribution of the protein system described above and (c) shows the communication schemes derived from an imbalanced load distribution.}
\label{fig: hespadda}
\end{figure}

Simulating heterogeneous molecular systems on supercomputers requires the conception and development of efficient parallelization techniques or Domain Decomposition (DD) schemes~\cite{lammps95,gromacs4,de_shaw}. In the first release of ESPResSo++ the Domain Decomposition scheme was a combination of the Linked-Cell-List algorithm (LCL) with an homogeneous-spatial Domain Decomposition. Such schemes are applied to traditional molecular simulations, for instance in dense homogeneous polymer melt systems~\cite{Kurt90,moreira2015direct,zhang2014equilibration}. 
While traditional molecular simulations are performed with the same resolution for all molecules in the simulation box; in heterogeneous systems~\cite{GuzmanPRE2017}, we tackle different resolutions (densities). Spatially the simulation box is typically comprised by subregions with different resolutions, namely, for multiscale simulations the coarse-grained and the atomistic/hybrid subregions (see Figure~\ref{fig: hespadda}(a)). In terms of computational costs, the most expensive regions are the ones containing atomistic details (higher resolution) followed by the regions using coarse-grained models~\cite{praprotnik_adaptive_2007-1,praprotnik_corrigendum:_2009} or ideal gas~\cite{Kreis2015} which are significantly cheaper.

The challenges faced by domain decomposition algorithms arise from two factors, namely, the interactions per subdomain and the communication between subdomains. An example of the inter-domains communication constraint is the imbalanced amount of data communicated between the fully atomistic and hybrid regions in comparison to the CG regions (see Figure~\ref{fig: hespadda}(c)). In addition, the interactions per domain will be imbalanced if an homogeneous grid decomposes equally the system as shown in Figure~\ref{fig: hespadda}(b). This is mainly because the distribution of interactions per subdomain could differ in orders of magnitude depending on the spatial heterogeneous mapping ratio as defined in Equation~\ref{eq:R_MS}.

\begin{equation}\label{eq:R_MS}
R^{res}_{SH}=\frac{N^{res}_{HR}}{N^{res}_{LR}}
\end{equation}

where $N^{res}_{HR}$ is the number of entities in the high-resolution region that corresponds to one entity in the low-resolution one $N^{res}_{LR}$. For example mapping the atomistic water molecule to the coarse-grained model can usually result in a $R_{SH}^{res}=3$~\cite{GuzmanPRE2017} (See also Figure~\ref{fig: hespadda2}(a)).

To tackle the aforementioned limitations (illustrated in Figures~\ref{fig: hespadda}(b) and ~\ref{fig: hespadda}(c)), the updated release of ESPResSo++ includes an implementation of the Heterogeneous Spatial Domain Decomposition Algorithm, for short HeSpaDDA~\cite{GuzmanPRE2017}. In a nutshell, HeSpaDDA will make use of \textit{a priori} knowledge of the system setup, meaning the region that is computationally less expensive. This inherent load-imbalance could come from different resolutions or different densities. The algorithm will then propose a non-uniform domain layout, \emph{i.e.} domains of different size and their distribution amongst compute instances. This can lead to significant speedups for systems of the aforementioned type over standard algorithms, \emph{e.g.} spatial Domain Decomposition~\cite{lammps95} or spatial and force based DD~\cite{de_shaw}.

\subsubsection*{Algorithm description}
\begin{figure}[thb!]
\centering
\includegraphics[clip,width=1.0\columnwidth,keepaspectratio]{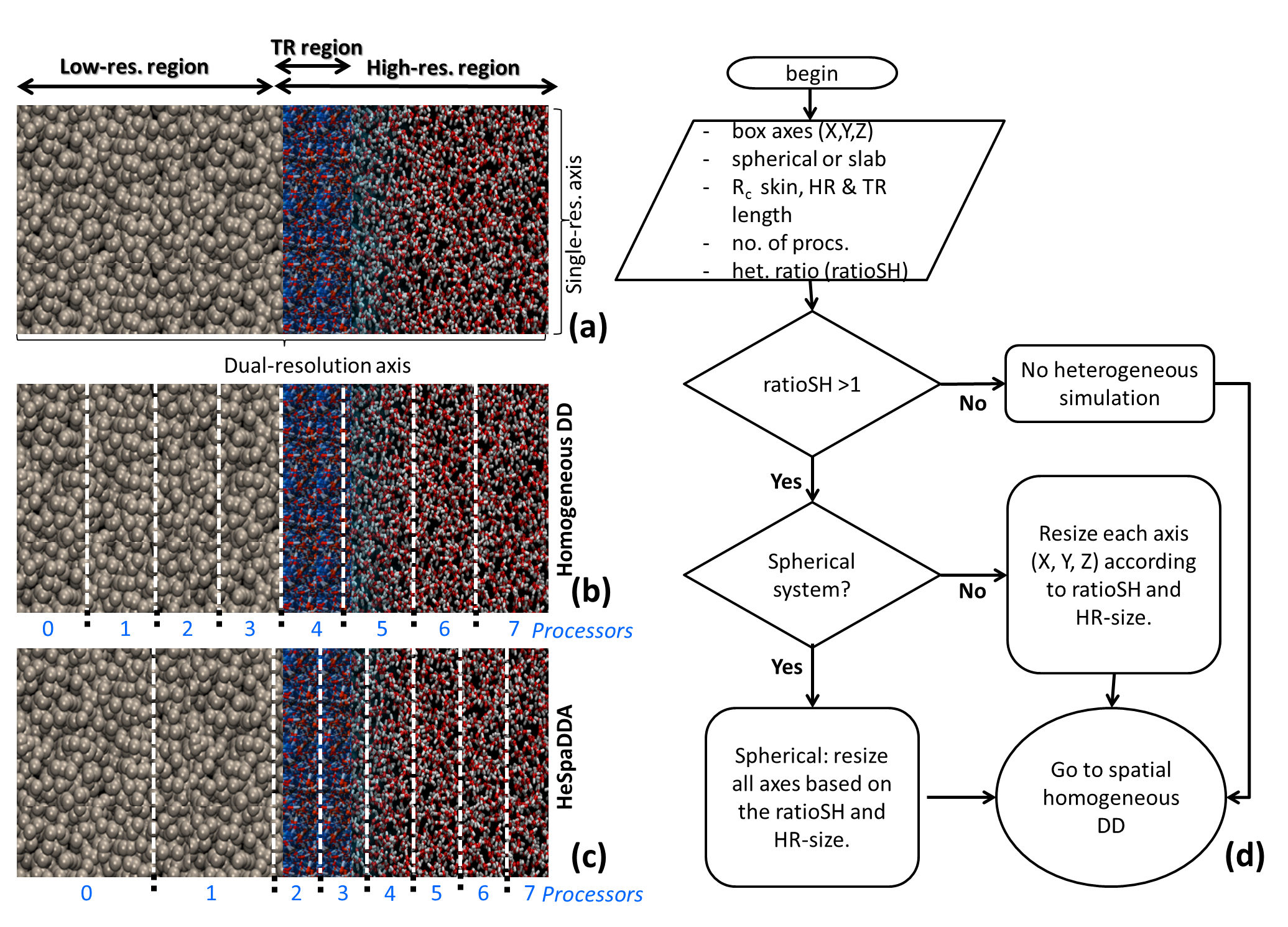}
\caption{AdResS simulation of water (a) coarse-grained (low-res. region), Hybrid (TR-region) and the Atomistic regions are illustrated. The two latter make up the high-res. region. (b) depicts the processor allocation for the homogeneous one-dimensional domain decomposition, while (c) for HeSpaDDA. (d) shows the processors allocation flowchart of HeSpaDDA.}
\label{fig: hespadda2}
\end{figure}

The proper allocation of processors in heterogeneous molecular simulations is vital for the intrinsic computational scaling and performance of the production run. Moreover, the whole simulation performance is constrained to the initial domain decomposition and hence the correspondence of the number of processors to different resolution regions of the initial given configuration. An example of such a heterogeneous initial domain decomposition is provided by a multiscale simulation of water, where the system is decomposed in the x-axis by 8 processors and the homogeneous and HeSpaDDA cases are depicted one next to the other(see Figures~\ref{fig: hespadda2}(b) and~\ref{fig: hespadda2} (c)). Also the algorithm flowchart can be found in Figure~\ref{fig: hespadda2}(d).

Once the processors allocation has been built, an initial cells distribution per subdomain is created to find the maximum number of cells to be used per region. Such a distribution can be done either symmetrically and non-symmetrically. The symmetric cells distribution is triggered if the heterogeneous regions can be decomposed as mirrors within half of the box and the non-symmetric distribution occurs if the heterogeneous regions cannot be mirrored within the simulation box. Within those functions, control statements check if the number of processors to be used are even or odd, as well as the number of cells in each dimension. In case there are still non-distributed cells the symmetric and non-symmetric functions will call a pseudo-random weighted cells distribution for the remaining ones. As a final step the algorithm verifies if the performed DD is scalable, and suggests a possible number of cores to perform the heterogeneous simulation (a Python function named cherry picked processors \textit{cherrypickTotalProcs}). A detailed description of the processors allocation and cells distribution algorithms are provided in a previous article~\cite{GuzmanPRE2017} and all python scripts can be found in the referenced code~\cite{eppGH}.

\subsubsection*{Implementation in ESPResSo++}
The implementation of HeSpaDDA in ESPResSo++ has involved the creation of new data structures, for the number of cells inhomogeneously distributed in each subdomain as depicted in Figure~\ref{fig: hespadda2}(C). Such data structures are linked to an iterative algorithm that allocates processors to the simulation box according to the resolution of regions \emph{i.e.} fully atomistic, coarse-grained, among others. The processors allocation algorithm flowchart is described in Figure~\ref{fig: hespadda2}(d).

\section{Development workflow\label{sec:devWF}}

Since the last release~\cite{Halverson2013} we have moved to GitHub hosting and hence from Mercurial to Git~\href{https://git-scm.com/}{https://git-scm.com/} as a version control system.
We have also set a new development workflow, {\em fork-and-branch}, which is commonly used on the GitHub platform~\cite{Scott2015}.

This approach requires two things from the developers: fork the repository and use pull requests to ask for changes in the code base.
Basically, every new feature is developed on a branch of the developer's fork repository. 
Once a feature reaches completion, a merge request is sent to the default branch via GitHub's pull request mechanism. We use {\em master} branch as the default development branch.
The pull request is then reviewed by one of the ESPResSo++ core developers. Usually, minor improvements like e.g. adding tests or documentation are requested from the feature developer. Once all the newly added and existing tests pass, the feature is merged into the master branch.

This whole workflow is supported by continuous integration (CI) tests\cite{CI}, meaning before the pull request is accepted, it has to fulfill three conditions: properly build, pass all unit-tests and do not decrease code coverage.

The build process is pursued under three Linux distributions: Ubuntu (latest and long-term-support), Fedora, and OpenSUSE.
Moreover, every change is checked against two different compilers, gcc~\cite{Stallman2016gnu} (versions 4.7, 4.8, 4.9) and clang~\cite{Lattner2008llvm} as well as the internal and external Boost library~\cite{Karlsson2005beyond}.

As for the tests, we use two types, one that tests particular features, so-called unit-tests.
The second set, the regression tests, run against existing reference data.
CI gives the developers the advantage that even before the actual code review within the pull request it is easy to see if any changes broke existing tests. 

The last condition is the code coverage, which describes the percentage of code lines that were tested with unit-tests.
For that, we use the external tool (\href{https://codecov.io/}{https://codecov.io/}).
The current approximate code coverage, calculate by that tool, is 44\%.

If all conditions are matched then the pull request can be accepted.
In addition, we use CI to build different pieces of documentation including the website, the Doxygen~\cite{doxygen} documentation of the code and a pdf of the user's guide. The newly generated documentation is automatically deployed to \href{http://espressopp.github.io/}{http://espressopp.github.io/}. 
This documentation is available to the developer to support the usage of the most recent development version. 
This continuous deployment allows to not waste any time using the possibly outdated documentation of the last release.
Moreover, the latest master version is deployed and released to Docker Hub for users who prefer to test ESPResSo++ without building it themselves.

The release versions of ESPResSo++ follow the idea of semantic versioning~\cite{PrestonWerner} and the releases workflow ties into it very easily. 
In a nutshell, fixes and small new features that don't change the application interface can go into "stable" and hence only trigger a minor release, while big refactors that introduce a new feature or break backward compatibility of the application interface go into a major release.
After each major release, a {\em stable} branch is created and git tags on that branch mark the individual minor releases. Bugs and hotfixes are merged (via the same pull request workflow) into the stable branch directly. If necessary, the stable branch is merged into the master branch occasionally to include all fixes into the master branch as well.
Here, we present version 2.0 of the package.

Moreover, in the spirit of reproducibility, we have deposited the release on the zenodo.org respository\cite{espp_zenodo}.

\section{Integration with other packages\label{sec:intO}}

The ESPResSo++ can easily be used with other software packages, primarily due to its internal design that allows it to work as nothing more than a Python module.
In this way, ESPResSo++ can be called from any other Python code, even from Jupyter~\cite{Perez2007} during an interactive session.
This allows using ESPResSo++ as an educational tool during hands-on sessions.

The recent release of ESPResSo++ brings support for the new file format H5MD~\cite{DeBuyl2014} that uses HDF5~\cite{hdf5} storage.
The H5MD file format is suitable to hold information about the simulation details in a self-descriptive, binary and portable format.
Along with the particle positions, with this file format it is possible to store information of box size, particle types, mass, partial charges, velocity, forces and other properties like software version, integrator time-step and a seed of random number generator.
Together with the information about particles, the H5MD can store information about the connectivity, in a static or dynamic manner.
The static information comprises bonds, angles and dihedrals which are stored at the beginning of the simulation and they would not be updated during the run.
Moreover, a dynamic information storage can track the changes in the bonds, angles and dihedrals also during the simulation.
This can be very important, e.g, when we use ESPReSo++ to perform chemical reactions~\cite{DeBuyl2015,Krajniak2018}.
By this, it is possible to share not only results of the simulation but also certain details that allow reproducing these results.
In addition, HDF5 storage natively supports parallel input-output (I/O) operations which allows performing efficient parallel simulations.

Apart from the H5MD file format, ESPResSo++ has trajectory writers to GROMACS~\cite{Hess2008}, XTC, XYZ and PDB file formats.
The simulation details can also be stored in LAMMPS~\cite{lammps95} file format.
When it comes to reading, the GROMACS topology and trajectory file formats are supported, hence it is possible to run directly a simulation from those input files.
This is also true of LAMMPS input files. 
Because of the variety of file formats that are supported by ESPResSo++, the integration with existing packages is very easy, for example, VOTCA 1.3~\cite{votca13} package can cooperate with ESPResSo++ out-of-the-box.

\section{Examples and documentation}

All features and their implementation are described in detail in the ESPResSo++ documentation~\cite{eppdoc}. Furthermore, ESPResSo++ comes with many example scripts~\cite{eppGHexamples} and tutorials that cover all methodologies and demonstrate to the user how to set up various types of simulation systems in practice.

\section{Conclusions}
In this article we have presented several state-of-the-art multiscale molecular simulation methods that have recently been implemented in the ESPResSo++ package. Specifically, we discussed the latest advancements of adaptive resolution simulations as well as another multiscale scheme for coupling lattice Boltzmann and molecular dynamics techniques. Furthermore, a hierarchical strategy for equilibrating polymer melts and an accompanying domain decomposition scheme, which have also been implemented into ESPResSo++, were presented. The deployment of those methods and algorithms shows the flexibility and extensibility offered by the software package for method development of advanced molecular simulations techniques.

From the software development viewpoint, we are providing scientists from the material and biomolecular scientific communities a computational tool that can be used out-of-the-box from simple Python scripts and allows exploring diverse molecular systems in polymer research, membranes, proteins, crystallization processes, evaporation, among others. This update is also very useful for the prototyping of new theoretical concepts and molecular simulation method development since it includes proven functionalities like: (i) extension of new simulation methods on top of the presented ones, as shown in Section~\ref{adress}, (ii) the development of new algorithms in Section~\ref{hierarchical} and Section~\ref{mrDD}, or (iii) the combination of multiple scales and multiple methods like the development shown in Section~\ref{latB}. 

We have covered many possible applications of the recent methods and algorithms included within the updated ESPResSo++, in particular for the areas of soft matter science, as illustrated within each section of this article. Selected applications of the new ESPResSo++ release have been published and are also referenced within this release communication. We note that these improvements make the simulation of many different systems feasible, from short proteins to huge polymer melts, passing through advanced path integral multiscale systems.

On top of the aspects described above, the ESPResSo++ package is open source (published under the GNU General Public License (GPL) version 3) and hence offers the molecular simulation community the possibility of extending the package and/or adapting methods to their research interests. Moreover a friendly developers environment, including recent developers software tools, mailing lists, repository management, an improved documentation and even parsing of input files from other MD packages, aims to smooth the transition from such packages to ESPResSo++.

\section{Acknowledgments}
T. Stuehn, N. Tretyakov and H. V. Guzman acknowledge financial support under the project SFB-TRR146 of the Deutsche Forschungsgemeinschaft. H. Kobayashi acknowledges the European Union’s Horizon 2020 research and innovation program under the grant agreement No. 676531 (project E-CAM). A.C. Fogarty acknowledges research funding through the European Research Council under the
European Union’s Seventh Framework Programme (FP7/2007-2
013) / ERC grant agreement no.
340906-MOLPROCOMP. 
 J. Krajniak acknowledges ``Strategic Initiative Materials'' in Flanders (SIM) under the InterPoCo program and VSC (Flemish Supercomputer Center; Hercules Foundation and the Flemish Government - department EWI).
This work has been partially authored by an employee of Los Alamos National Security, LLC, operator of the Los Alamos National Laboratory under Contract No. DE-AC52-06NA25396 with the U.S. Department of Energy. The United States Government retains and the publisher, by accepting this work for publication, acknowledges that the United States Government retains a nonexclusive, paid-up, irrevocable, world-wide license to publish or reproduce this work, or allow others to do so for United States Government purposes. C. J. thanks Los Alamos National Laboratory for a Director’s Postdoctoral fellowship supporting the early stage of this work. Assigned: LA-UR-18-XXXXX.

\section{References}
\bibliographystyle{cpc}
\bibliography{pubs,pubs_nik,pubs_hk}

\end{document}